 \documentclass[preprint2]{aastex}
 \usepackage{multirow}
 \usepackage{epstopdf}

\shorttitle{Influence of initial mass segregation on runaway collisions}
\shortauthors{Goswami, et al.}
\begin{document}

\title{Formation of Massive Black Holes in Dense Star Clusters. II. IMF and Primordial Mass Segregation}

\author{Sanghamitra Goswami,\altaffilmark{1} Stefan Umbreit,\altaffilmark{1} Matt Bierbaum,\altaffilmark{2} Frederic A.\ Rasio\altaffilmark{1,3}}
\affil{$^{1}$ Department of Physics and Astronomy, Dearborn University, Northwestern University, 
Evanston, IL 60208, USA}
\affil{$^{2}$ Department of Physics, Clark Hall, Cornell University, Ithaca, NY 14853, USA}
\affil{$^{3}$ Center for Interdisciplinary Exploration and Research in Astrophysics (CIERA), Northwestern University.}





\begin{abstract} 
  A promising mechanism to form intermediate-mass black holes
  (IMBHs) is the runaway merger in dense star clusters, where main-sequence stars collide and
  form a very massive star (VMS), which then collapses to a black hole. In this
  paper we study the effects of primordial mass segregation and the importance of the stellar initial mass function (IMF) on the runaway
  growth of VMSs using a dynamical Monte Carlo code for $N$-body systems with $N$
  as high as $10^{6}$ stars. Our code now includes an explicit treatment of all stellar collisions. We place special
  emphasis on the possibility of top-heavy IMFs, as observed in some very young massive clusters.  We find that both primordial mass segregation and the shape of the IMF affect the rate of core collapse of star clusters and thus the time of the runaway. When we include primordial mass segregation we generally see a decrease in core collapse time ($t_{\rm cc}$). Although for smaller degrees of primordial mass segregation this decrease in $t_{\rm cc}$ is mostly due to the change in the density profile of the cluster, for highly mass-segregated (primordial) clusters, it is the increase in the average mass in the core which reduces central relaxation time decreasing $t_{\rm cc}$. The final mass of the VMS formed is always close to $\sim10^{-3}$ of the total cluster mass, in agreement with the previous studies and is reminiscent of the observed correlation between the central black hole mass and the bulge mass of the galaxies. As the degree of primordial mass segregation is increased, the mass of the VMS increases at most by a factor of $3$.
 Flatter IMFs generally increase the average mass in the whole cluster,  which increases $t_{\rm cc}$. For the range of IMFs investigated in this paper, this increase in $t_{\rm cc}$ is to some degree balanced by stellar collisions, which accelerate core collapse. Thus there is no significant change in $t_{\rm cc}$ for the somewhat flatter global IMFs observed 
  in very young massive clusters.
  \end{abstract}



\section{Introduction}

\subsection{IMBHs} 

It is generally accepted that there exist two separate classes of black holes (BHs)
defined by their mass ranges: stellar-mass BHs of mass
$\sim3 - 20\, {\rm M_{\odot}}$ \citep{Bolton72, WebMur72, Casares07}  that form
through the collapse of massive stars, and supermassive BHs (SMBHs) of
mass $\sim10^6 - 10^{10}\, {\rm M_{\odot}}$ that are found in the centers of most
galaxies including the Milky Way \citep{korm, merritt01, gz2003, miller04}.
However, this leaves a gap in the mass range from $\sim10^2 - 10^4\, {\rm
M_{\odot}}$ which in recent years seems to have been filled by tentative
evidence for intermediate-mass black holes (IMBHs) \citep{zwart99, Grin01}.
This evidence consists of dynamical measurements \citep{Gebhardt02, gerssen02,
Van02}, detection of ultraluminous X-ray sources (ULX) found off-center in galaxies \citep{miller04}, the anomalously high
mass-to-light ratio observed in the centers of some globular clusters
\citep{miller04} and the mass segregation quenching in the cores of globular clusters in presence of an IMBH \citep{Pas09}.

The possible existence of IMBHs in globular clusters is suggested by the $M_{\rm BH}-\sigma$ relation for galaxies, where $M_{\rm BH}$ is the mass of
the central massive BH, $\sigma$ the velocity dispersion of the
bulge. Extending this relation down to velocity dispersions typical for
globular clusters ($\approx 10\,\rm{km\,s}^{-1}$), we expect BH masses
in the range of $\sim10^3 - 10^4\,\rm M_{\odot}$. There are several claims in the
literature that, indeed, such IMBHs are present in some globular clusters. The
evidence is mainly based on measurements of the stellar velocity distribution,
with velocity dispersions increasing strongly towards the center if an IMBH is
present. The most promising candidates are the clusters G1
\citep{Gebhardt05,Gebhardt02} and $\omega$ Cen \citep{noyola2008}. However, the
presence of IMBHs in these clusters is still debated. For instance, based on
scaled direct $N$-body models \citet{Baumgardt03} show that there is no need to
invoke the presence of an IMBH in the center of G1 in order to explain the
observed velocity dispersion profile. On the other hand, \citet{Gebhardt05}
point out that, since the region where the IMBH influences the velocity
distribution is barely resolved, additional information from higher order
velocity moments must be considered. Using orbit-based models to fit both
surface brightness and velocity data simultaneously, they find, similar to
their earlier results, that only these higher order velocity moments provide
sufficient evidence for an IMBH in G1 with a mass of $1.8(\pm 0.5)\times
10^{4}\,\rm{M}_{\odot}$. However, these results must still be confirmed using
fully self-consistent evolutionary models that do not rely on assumed
mass-to-light ratio profiles.

Numerous ULX sources have been identified by Chandra and XMM-Newton, often
associated with starburst environments. These sources have
X-ray luminosities of $L_X > 10^{39}\,{\rm ergs^{-1}}$, exceeding the
angle-averaged flux of a BH of mass $< 20\,{\rm M_{\odot}}$ accreting material
at the Eddington luminosity, $L_E$. Although many ULX could be identified as
active galactic nuclei at the centers of galaxies, and thus SMBHs, some are
clearly off-center with typical projected distances of $\approx 400\,\rm pc$
\citep{miller04}. For instance, the ULX X41.4+60 
in the starburst galaxy M82
\citep{P04} has been found 7" away from the galaxy center which corresponds to
a distance of $\approx 200\,\rm{pc}$. \citet{Kaaret01} argue, therefore, that
this X-ray source is unlikely to be an under-luminous SMBH, as dynamical
friction would cause it to spiral into the nucleus of the galaxy on timescales
much shorter than the age of the galaxy.  Based on this argument they derive an
upper mass limit of $\approx 10^5\,\rm{M_{\odot}}$. From the total X-ray
luminosity and the assumtion that the ULX source radiates isotropically at the
Eddington rate, \citet{Kaaret01} also derive a lower limit of
$500\,\rm{M}_{\odot}$. However, they point out that this value can be lower by
a factor of a few when mild beaming is considered, thus bringing it closer to
the stellar mass BH range. On the other hand, newer Chandra observations by
\citet{Kaaret09} have shown that during an outburst, where X41.4+60 increased
its X-ray luminosity by a factor of more than $3$, its X-ray spectrum remained
in the so called ``hard state", characterized by a dominant power-law component
containing $80\%$ of the total flux \citep{REMC}. From stellar-mass BH X-ray binaries it is known that
their spectra are in the hard state if $L<0.3\, L_E$. If one assumes the same
limit for higher-mass BH X-ray binaries, the peak luminosity of $8.5\times
10^{40}\,\rm{ergs}^{-1}$ inferred from Chandra observations would imply a
lower mass limit of $\approx 2000\,\rm{M}_{\odot}$ \citep{Kaaret09}. As one can
see, even if one considers that the radiation could be mildly beamed, the
minimum mass is well above the stellar-mass BH range, making this source a
prime IMBH candidate. Another good candidate for an IMBH is the recently found
ULX source HLX-1 in the edge-on spiral galaxy ESO 243-49 \citep{farrell09}.
With a maximum luminosity of up to $1.1\times10^{42}\,{\rm ergs^{-1}}$ in the
$0.2$-$10\, {\rm kev}$ band, and accounting for beaming effects,
\citet{farrell09} obtain a conservative lower limit of $\sim 500\,{\rm
M_{\odot}}$.

Further indications for the presence of IMBHs in some globular clusters come
from unusually high mass-to-light ratios measured in their centers, as inferred
from pulsar timing measurements. For instance, the Galactic globular cluster
NGC 6752 has five millisecond pulsars; three of which are in
the core \citep{damico02}. Two of these three have negative period derivatives
and one has an anomalously high positive period derivative. If the spin
derivatives are due to the gravitational potential of the cluster,
\citet{ferraro03} conclude that the mass-to-light ratio in the core is M/L $\approx 6 - 7$, much higher than inferred for most globular
clusters. For comparison, $\rm M/L\approx 2$ is what one would expect for an old
star cluster with standard IMF based on stellar evolution alone
\citep{caputo85}. Furthermore, based on the position of one of the pulsars,
\citet{ferraro03} find that there is $1000 - 2000\, \rm M_{\odot}$ of
underluminous matter within the inner $0.08\, \rm pc$ of the cluster.  This could
be explained by a $\sim 1000\, \rm M_{\odot}$ IMBH in the center of the cluster,
but also by an exceptional concentration of dark remnants, or a $\sim 100\, \rm
M_{\odot}$ black hole that is offset but near the projected location of the
three millisecond pulsars in the cluster core. As the high spatial
resolution in their observations does not show a
cusp down to $0.08\, \rm pc$, \citet{ferraro03} conclude that any central BH must have a mass $\rm M \lesssim 1000\, \rm M_{\odot}$.

There has been some considerable theoretical work by \citet{gil08} on observational evidences of IMBHs in star clusters. Mass segregation in star clusters, brings heavier stars towards the centre increasing the average stellar mass in the core. As a diagnostic tool to quantify mass segregation, \citet{gil08} defined this radial variation of the average stellar mass as $\Delta m=\left<m\right>_c-\left<m\right>_{\rm rh}$, where $\Delta m$ is the difference of the average mass in the core ($\left<m\right>_c$) and the average mass at the projected half mass radius of the cluster ($\left<m\right>_{\rm rh}$). A cluster with no IMBH, on a relaxation timescale settles to a quasi-equilibrium configuration with varying degrees of mass segregation ($\Delta m$). They found that in simulations of clusters with an IMBH, mass-segregation ($\Delta m$) is significantly quenched.
The idea is that, if there is an IMBH in the core, since the IMBH will be more massive 
than any of the other massive stars in the core, the IMBH has an extremely high probability of exchanging into a binary in a close three-body encounter. The subsequent interactions of this IMBH in a binary, with other massive stars in the core might kick out or scatter the other massive stars in the core. Since it will be the mass-segregated massive stars which will be scattered out of the core in this way, $\left<m\right>_c$ will decrease. 
Thus, in a cluster with no central IMBH, mass segregation ($\Delta m$) will be more pronounced than a cluster with an IMBH. According to the authors this phenomenon of quenching of mass segregation in presence of an IMBH can be observed by high-resolution imaging of the of the cores of a large sample of globular clusters by Hubble Space Telescope \citep{Pas09}. 


\subsection{Pathways to IMBH formation}

There are several pathways discussed in the literature through which IMBHs may
form. The simplest way is through the core collapse of a massive Population III
star formed in a mini dark matter halo at high redshift \citep{Madau01}. At
lower redshifts such massive stars must be grown through mergers of lower mass
stars, which requires rather large stellar densities, typically 
$\geq 10^6\, \rm{pc}^{-3}$ \citep{ freitag06, ardi08,  Baum08}. 
As an alternative to stellar mergers it is also possible to increase the mass of 
a stellar mass BH by tidally disrupting and accreting other stars;
 however, this seems to require even larger stellar densities \citep{Baum08}. 
 
In a cluster with a broad range of stellar masses, large stellar densities can be
achieved through mass segregation, which will drive the most massive objects to
the center, while lighter stars spread out to attain kinetic energy
equipartition. However, for any reasonable IMF the most massive stars are
unable to achieve energy equipartition with the lighter stars and therefore
decouple from the rest of the cluster and form a compact subsystem in the
cluster center. This process is called Spitzer instability \citep{spit1969} and
causes core collapse to be accelerated \citep{spit1969, Vish78, watters00,
GU04}. As shown by \citet[][hereafter Paper I]{GU04}, for a realistically broad stellar mass
spectrum the core collapse time ($t_{\rm cc}$) can be as short as $t_{\rm cc} = 0.15 t_{\rm
rc}(0)$, where $t_{\rm rc}(0)$ is the initial central relaxation time.

Once the most massive
objects have segregated, the formation of an IMBH can occur in two different
ways: one is when the massive stars have already evolved into stellar-mass BHs
at the time of core collapse and these BHs then merge by emitting gravitational
waves. The BHs first form BH binaries which then harden through 
dynamical interactions with other objects until the binary is close enough
for gravitational radiation to dissipate sufficient orbital energy until the
BHs merge \citep[see, e.g.,][]{OL06}. However, growing an IMBH with a mass of
$\sim 1000\,{\rm M}_{\odot}$ through such stellar-mass BH mergers is only
realistic for the very massive clusters such as those in galactic nuclei \citep{OL06}. For
smaller systems such as globular clusters \citet{MOTA02} and \citet{OL06} have shown that this mechanism
is rather inefficient. This is because it is much more likely that the stellar mass
BHs escape through strong dynamical binary interactions \citep{PM00} or recoil from
asymmetric gravitational wave emission \citep{MIHA02, GUL04, OL07} given the
low cluster escape speed.


Another way to form an IMBH, which is not restricted to galactic nuclei,
is through mergers of massive main-sequence stars that segregate to the center
and drive cluster core collapse before formation of stellar
BHs. This subsystem of massive stars can enter a phase of rapid collisions, and
since the most massive object has the largest cross-section for further
collisions, this object is expected to grow
in a runaway fashion. The resulting VMS eventually collapses to form an IMBH \citep{PM02,
GU04}. As the time it takes for the most massive stars to turn into BHs is
approximately $3\,{\rm Myr}$ \citep{MEMA00}, an IMBH is only formed through runaway
merging if the cluster reaches core collapse within the first $3\,{\rm Myr}$ of
dynamical evolution.

 This simple picture has a few important caveats.
First, the fate of such a massive merger remnant formed by a runaway is rather uncertain. Direct monolithic collapse to a BH with no or little mass loss is a
possible outcome, at least for sufficiently small metallicities \citep{HE02}. However, it has been suggested that mass
loss from stellar winds could dominate the mass increase due to repeated
mergers \citep{GL09}. In this case it might be difficult to
form a VMS at all. On the other hand,
within the runaway phase, it has been shown that, for clusters like those studied here, the
time between collisions is much shorter
than the Kelvin-Helmholtz timescale of the collision product, so that the growing VMS
must be out of thermal equilibrium.
Instead the calculations of \citet{GL09} were done assuming that each merger remnant behaves exactly like an ordinary massive star in thermal equilibrium.
The uncertainty in the final VMS (and IMBH) mass is further increased when considering
that a stronger wind mass loss would also lead to a stronger expansion of the cluster
core, decreasing the mass growth rate of the VMS by lowering the collision rate. To address this problem, one would need to perform a fully self-consistent simulation coupling the stellar
dynamics with detail radiation hydrodynamics of the stellar collisions and mass loss from merger remnants. This is clearly beyond the scope of this paper.

The runaway collision scenario in globular clusters has been extensively
investigated in Paper I and in \citet{freitag06,freitagrasio06} using Monte Carlo (MC)
simulations for a large variety of initial conditions.  Paper I focuses on the
dependence of $t_{\rm cc}$ on the shape and the width of the IMF, the presence
of a Galactic tidal field, and the cluster density profile.  The key
result is that for clusters with a broad range of masses, $t_{\rm cc}$ is set by
the central relaxation time, $t_{\rm rc}(0)$, which means that for multi-mass
clusters core collapse depends on the local conditions in the core while for
single-mass clusters core collapse is a global phenomenon. Furthermore it is clear that $t_{\rm cc}$ depends only weakly on the external tidal field. It was also found that the dependence of $t_{\rm cc}$ on the mass spectrum can be
conveniently expressed by a single parameter $m_{max}/\left<m\right>$, where
$m_{max}$ is the maximum and $\left<m\right>$ the average stellar mass.  For
$m_{max}/\left<m\right>\ > 50$ the ratio $t_{\rm cc}/t_{\rm rc}$ converges to a
constant value $\approx$ $0.15$ for all IMFs and cluster density profiles. Combined
with the requirement that $t_{\rm cc}$ be less than $3\,\rm Myr$, this relation
provides a uniform criterion for runaway growth to occur for a large range of
possible, unsegregated cluster configurations. 


\citet{freitag06} performed similar simulations but also incorporating
collisions explicitly. They quantified the dependence of the onset of the
runaway on the initial collision time, and found that runaway growth happens
earlier with respect to $t_{\rm rh}$, the half-mass relaxation time for
initially more collisional clusters. Thus, collisions extend the parameter
space of initial cluster conditions for runaway to occur.  However, as pointed
out by the authors, for any standard IMF (Kroupa, Salpeter, Miller-Scalo) this
effect is negligible for masses typical for globular clusters. Thus, in this
regime, the condition for runaway to occur reduces to the one found in Paper I, based on the central relaxation time alone. 

The runaway collision scenario has also been verified numerically by direct
$N$-body simulations. In \citet{PM02} runaway collisions
were produced in sufficiently dense and highly concentrated clusters with
only a few $10^4$ stars initially. \citet{P04}  modelled the
evolution of MGG-11 with $\approx 10^5$ and, in one case, with $\approx
5\times10^5$ stars, and found that, similar to Paper I, only
clusters with a short enough mass-segregation timescale, or, correspondingly,
low enough $t_{\rm rh}$, are likely to produce a runaway object. However, their
results also show that, in addition to a short $t_{\rm rh}$, these clusters must
also be sufficiently concentrated, corresponding to King models with
${\rm W_0}\geq9$, in order to trigger a runaway object. This might be due to the fact
that in the $N$-body simulations binaries formed by three-body interactions
\citep{freitag06}, a process which is not included in the MC
runs, and these binaries dominated the collisional evolution in the core.  On
the other hand, for $N\gtrsim10^6$ \citet{freitag06} have demonstrated
that three-body binaries are of little importance for the collision process.
Similarly, \citet{P04} find for their large
$N\approx5\times10^5$ run, that the influence of three-body binaries on the
collisional evolution is lower than in their lower-$N$ runs. Thus, it appears
that the nature of collisional runaway growth in a star cluster changes
with increasing $N$ from being less dominated by binary collisions and
involving more and more single-single collisions.

Although, the runaway merger scenario has been extensively investigated for a
large variety of initial conditions, almost all of these studies started with
unsegregated clusters, that is, clusters, where the stellar mass is not
correlated with the radial position within the cluster. However, in recent
years observations frequently indicated that many young star clusters with
ages of only a few Myr show already a strong degree of mass segregation,
suggesting that mass segregation may be primordial and motivating a new study
of its effects on collisional runaways.

\subsection{Primordial Mass Segregation and Top-Heavy IMF} 

Significant mass segregation has been found in many young star clusters.
Mass segregation can be understood as the tendency towards equipartition
of 
kinetic energies. This tendency for equipartition is a consequence of gravitational encounters,
which attempt to drive the local velocity distribution toward a Maxwellian, with $m_1\left<V_1^2\right> = m_2\left<V_2^2\right>$ \citep[Chapter 16]{heghut03}.
As a result, massive stars move more slowly, on average, than lighter 
ones, so the massive stars drop lower in the potential well, while the stars of 
smaller mass move out and may even escape \citep[Chapter 16]{heghut03}.
This results in having the more massive members of a gravitationally bound 
system closer to the center whereas the lighter members are found further away.  
This dynamical mass segregation acts on a time-scale of the order of the 
relaxation time of the cluster \citep[Section 4.2]{heghut03, Spi1987}.  


However, there are indications that the degree of mass segregation seen in many 
young clusters cannot be a result of their dynamical evolution as their ages 
are much less than their relaxation time \citep{HIllen97, Fisc98, Grijs98, 
HIHA98, Gouli04, Stolte06}.  For instance, for NGC330, the richest young star 
cluster in the small Magellanic Cloud (SMC), observations show that the IMF 
becomes steeper at increasing distances from the cluster center, with the 
number of massive stars decreasing from the core to the outskirts of the 
cluster 5 times more rapidly than the less-massive objects, while the age of 
NGC330 is 10 times shorter than the expected relaxation time of the cluster 
\citep{Siri02}. Another example is provided by the Orion nebula Cluster, in which it can be argued from numerical results that the massive stars in the centre cannot have formed in the outer regions. This implies that the stellar mass is to some degree a function of the initial position within the cluster \citep{BODA98}. Interestingly, the $3\,\rm mm$ continuum observations obtained with 
the Combined Array for Research in Millimeter-Wave Astronomy (CARMA) of 11 
Infrared Dark Cloud (IRDC) cores establishes that  mass segregation can be 
identified at the formative stage of a stellar cluster \citep{MULACA07}. All 
these observations suggest that many clusters may have been formed 
significantly mass segregated. 

Primordial mass segregation has been explained either by the formation of 
massive stars preferentially in the densest regions of the parent molecular 
cloud \citep{MULI96} or by competitive gas accretion during the earliest phases 
of star formation \citep{BOBA06}. An additional alternative has been suggested 
by \citet{MCVEPO07}. They show that a much higher degree of mass segregation in 
a cluster can be achieved if the cluster is a result of one or more mergers of 
smaller clusters. As the smaller clusters have shorter  relaxation times, mass 
segregation proceeds faster. When the clusters merge, this larger degree of 
mass segregation is preserved in the final cluster. 

Primordial mass segregation also has important consequences. It has been shown that, in an initially mass-segregated cluster, the 
effect of early mass loss due to stellar evolution is, in general, more 
destructive than for an unsegregated cluster with the same density profile, since this 
leads to shorter lifetimes, a faster initial evolution toward less concentrated 
structure and a flattening of the stellar mass function \citep{MCVEPO08}. 


Recently, \citet{ardi08} studied the influence of initial mass segregation
on the runaway growth of a massive star by means of direct $N$-body simulations of up to $\approx 131$ stars. Contrary to the expectations from
\citet{GU04}, they found that, for a given density profile, initial mass
segregation does not increase the available parameter space of cluster initial
conditions leading to runaway growth. They argue that this is because
initial mass segregation $decreases$ the collision rate of stars in the core as,
due to the increased average mass, the number density is decreased. This is in
line with earlier $N$-body simulations suggesting that runaway growth can occur only when the cluster
is initially sufficiently collisional, \citep{P04}, 
in contrast to predictions based only on the core collapse time. This result should be tested for larger $N$ as the nature of the collisional runaway changes from being dominated by three-body binary formation at low $N$ to single-single collisions at high $N$.


One of the main uncertainties in star
cluster evolution lies in determining the true initial mass function
(IMF). 
Often it is assumed that the IMF is a standard power-law (or power-laws with
different indices in different mass ranges) with no primordial radial variation
in the cluster \citep[e.g.,
][]{Sal1959,MS1979,Kr2001}.   
Deviations from the standard IMFs are observed in many clusters both at the high and low-mass ends \citep[e.g., ][]{Elme2004}.  In
particular at the high mass end MFs are observed to be generally flatter compared
to standard Salpeter power-law in young massive clusters like the Arches cluster
\citep{Stolte2002,kim06}.




\section{Numerical Methods}

\subsection{Monte Carlo code}

The numerical method that has been used here to investigate the dynamical evolution of star clusters is the Monte Carlo method, based on the classic work of \citet{Henon} and described in detail in \citet[][and references therein]{FreRas07}. 
In Monte Carlo simulations, $N$, the total number of stars in the cluster is dependent on the initial half mass relaxation time in the cluster. For a Plummer sphere it is given by \citep[eq. 2.63]{Spit87},
\begin{equation}
t_{\rm rh}(0)=\frac{0.138N}{ln\gamma_{c}N}\left(\frac{r_h^3}{GM}\right)^{1/2},
\end{equation}
where $\gamma_c\sim0.01$ is the coulomb logarithm and $r_h$ is the half mass radius.

Since our code now includes an explicit treatment of all stellar collisions
we briefly 
summarize here the `sticky sphere' method for stellar collisions \citep{FreBen02}.

In the sticky sphere approximation a collision occurs whenever the centers of two 
stars pass within a distance $d= (R_{1}+R_{2})$, with 
$R_{1,2}$ being the stellar radii.  Until this distance is reached, the 
gravitational influence of other stars as well as any mutual tidal interactions 
are neglected. The cross section for such a collision is given by \citet[section 7.5.8]{Bin87},
\begin{equation}
S^{12}_{coll}=\pi b_{max}^2= \pi (R_{1}+R_{2})^2(1+\frac{(v^{12}_{*})^2}{ v_{rel}^{2}}),
\end{equation}
where $b_{max}$ is the largest impact parameter leading to contact, $v_{rel}$ is the relative velocity between two stars and 
$v^{12}_{*} = (2G(M_{1}+M_{2})/(R_{1}+R_{2}))^{0.5}$. In a cluster where 
all stars have the same mass $M_{*}$ and radius $R_{*}$, the average local 
collision time $T_{coll}$ is given by \citet[section 7.5.8]{Bin87}

\begin{equation}
\frac{1}{T_{coll}} = \frac{1}{n_{*}} \int d^3v_1 d^3v_2f(v_{1})f(v_{2})||v_{1}-v_{2}||S_{coll},
\end{equation}
where $n_{*}$ is the stellar number density. For a Maxwellian velocity 
distribution this becomes 

\begin{equation}
  \frac{1}{T_{coll}}=16\sqrt{\pi}n_{*}\sigma_{v}R_{*}^{2}\left[1+\frac{GM_{*}}{2\sigma_{v}^{2}R_{*}}\right],
  \label{tcoll}
\end{equation}

where $\sigma_{v}$ is the velocity dispersion. 

In order to resolve collisional processes in a MC simulation we
constrain the time step size $\delta t$ according to an estimate of the central 
collision time using
\begin{equation}
  \delta t\leq f \tilde{T}_{coll},
\end{equation}
where $f= 5\times 10^{-3}$ is a constant chosen small enough to ensure that 
collisions are sampled sufficiently, and $\tilde{T}_{coll}$ is an estimate of 
$T_{coll}$ based on equation \ref{tcoll}, and given by
\begin{equation}
\frac{1} {\tilde{T}_{coll}}=16\sqrt{\pi}n_{*}\sigma_{v}\left\langle R_{*}^{2}\right\rangle \left[1+\frac{G\left\langle M_{*}R_{*}\right\rangle }{2\sigma_{v}^{2}\left\langle R_{*}^{2}\right\rangle }\right],
\end{equation}
with quantities in angular brackets being local averages. The collision probability for
two neighboring stars is calculated 
as
\begin{equation}
  P_{coll}=n_{*}v_{rel}S_{coll}\delta t,
\end{equation}
where $n_{*}$ is 
a local estimate of the stellar number density. The time step size $\delta t$, is chosen such that $P_{coll}<<1$. A random number (between $0$ and $1$) is then drawn. If the random number $< P_{coll}$ two stars are selected for a 
collision and they are merged under the assumption of mass and linear momentum conservation.



The units adopted for our simulations are the standard $N$-body units of the MC
code, and the same as in Paper I.

\subsection{Implementation Of Primordial Mass Segregation and Top-Heavy IMFs}
The effect of primordial mass segregation was studied in paper I
using a preliminary prescription. However, the recipe in Paper I is not 
defined with well constrained parameters. In this paper we have used recipes which are more realistically modelled and we do a much more extensive study.

Here we consider with 
two prescriptions for mass segregation: (i) the prescription of 
 \citet{Subr08} for generating mass-segregated clusters, in which the degree of mass segregation can be adjusted by a 
parameter $S$ related to the mean inter-particle energy of the 
stars; and (ii) the \citet{Baum08} prescription, which creates a maximally mass-segregated cluster in virial equilibrium.

From $N$-body models \citet{Subr08} find that the quantity that is 
transferred between the light and massive stars is the potential energy, while 
their average kinetic energy remains nearly constant during the cluster 
evolution. Therefore, mass segregation is generated in terms of mean inter-particle potentials $\left<U^{ij}\right>$, which is parameterized as
\begin{equation}
\label{mass-seg}
  \left<U^{ij}\right>=2(1-S)^{2}\left<U_{\rm tot}\right>\frac{M_{*i}M_{*j}}{M_{\rm tot}^{2}}\left(\frac{M_{\rm sub}^{ i}\, 
  M_{\rm sub}^{ j}}{M_{\rm tot}^{2}}\right)^{-S},
\end{equation}
where $M_{ *i}$ and $M_{ *j}$ denote the masses of the $i_{th}$ and $j_{th}$
particle, $M_{\rm tot}$ denotes the total mass of the cluster, $U_{\rm tot}$ is the total 
potential energy of the cluster, $M_{\rm sub}^{i,j}$ is the sum of all 
$M_{*k}<M_{*i,j}$ and $S$ is the degree of mass segregation which can have values 
between $0$ and $1$ (for a detailed description of the code refer to  \citet{Subr08}). In this recipe  $S=0$ implies an unsegregated cluster and $S=1$ a completely mass segregated cluster. Relating to 
entropy $S=0$ has the lowest entropy and the system is highly
symmetric in terms of $<U^{ij}>$, while for $S=1$, all the binding energy
is carried by the two most massive stars in the cluster and corresponds to a 
state of maximum entropy. With this parameterization of the inter-particle 
potential, quasi-stationary, star-by-star realizations of mass-segregated 
clusters are generated. The stars are assigned masses according to the
Salpeter IMF with maximum and minimum mass of $0.2\,\rm{M_{\odot}}$ and $120\,\rm{M_{\odot}}$ respectively.

An important property of this mass segregation recipe is that both the density 
profile and the average mass in the core ($\left<m\right>_{c}$)
is different for each value of $S$, starting from a Plummer sphere for 
$S=0$. This has the consequence that  $t_{\rm rc}(0)$ which is dependent on density in the core ($\rho_{c}$) and $\left<m\right>_{c}$  changes. This in turn causes $t_{\rm cc}$ to change.

The recipe from \citet{Baum08}, on the other hand, does not change the 
underlying density profile and generates mass-segregated clusters, in 
comparison, much faster. It essentially sorts all stars such that, for a given 
density profile, the most massive stars have, on average, the lowest specific 
total energy.

\begin{figure}[ht]
\begin{center}
\plotone{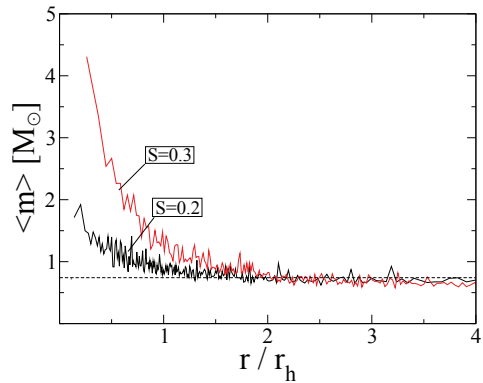}
\end{center}

\caption{\label{fig:Average-mass-profile2}Average stellar mass profile for initially
mass-segregated clusters created according to the prescription of
\citet{Subr08}. All the simulations shown here contain $5\times10^5$ stars with a Salpeter IMF (within $\rm M_{\rm min}=0.2\, \rm M_{\rm \odot}$ and $\rm M_{\rm max}=120\, \rm M_{\rm \odot}$). The horizontal axis shows the cluster radius in units of the initial half-mass radius. The dotted straight line is for an initially unsegregated cluster and the red and black solid straight lines are for clusters with different degrees of primordial mass segregation ($S=0.3, 0.2$). 
The average mass for the mass-segregated clusters increases steeply
within $\sim 1\, r_{\rm h}$. For $S=0.3$, the increase is by up to a factor
of $\sim5$.}

\end{figure}

\begin{figure}[ht]
\begin{center}
\plotone{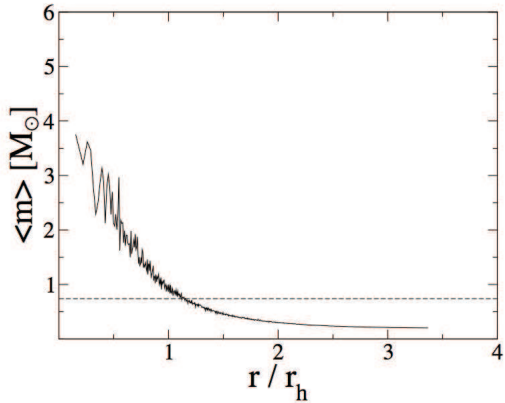}
\end{center}

\caption{\label{fig:Average-mass-profile1}Same as Fig.\ref{fig:Average-mass-profile2} but the primordially mass-segregated cluster is generated using the \citet{Baum08} recipe. Even in this case the average stellar mass of the cluster rises strongly within $1\, r_{\rm h}$, by up to a factor of $\sim
5$.}

\end{figure}

\begin{figure}[ht]
\begin{center}
\plotone{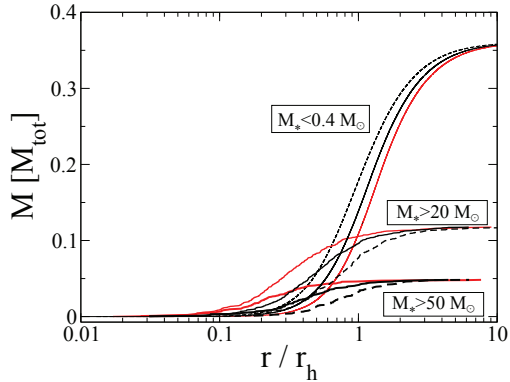}
\end{center}

\caption{\label{fig:Enclosed-mass2}Enclosed stellar mass in primordially mass-segregated (solid lines) and unsegregated (dashed lines) clusters. 
The average mass in the cluster is $\approx 0.68\, \rm M_{\odot}$. The black and red solid lines are for
mass-segregated clusters (\citet{Subr08} recipe) with the degree of mass segregation, $S = 0.2$ and $S = 0.3$, respectively. We show here mass enclosed in stars more massive than average mass of the cluster ($M_{*}>20\, \rm M_{\odot}$ and $M_{*}>50\, \rm M_{\odot}$) as well as mass enclosed in stars of mass less than average mass ($M_{*}<0.4\, \rm M_{\odot}$). All the simulations shown here contain $5\times10^5$ stars with a Salpeter IMF (within $\rm M_{\rm min}=0.2\, \rm M_{\rm \odot}$ and $\rm M_{\rm max}=120\, \rm M_{\rm \odot}$).   
Lines denoting enclosed stellar mass contained in stars more massive than the average mass (i.e., $M_{*}>20\, \rm M_{\odot}$ and $M_{*}>50\, \rm M_{\odot}$) show that the massive stars
in the unsegregated clusters are at significantly larger radii than the primordially mass-segregated cluster.  The
median position of massive stars in the unsegregated cluster differ by a factor of $\sim4$ for stars with $M_{*}>50\, \rm M_{\odot}$,
and by a factor of $\sim3$ for stars with $M_{*}>20\, \rm M_{\odot}$, compared to
the mass-segregated cluster with $S = 0.3$. Whereas lines denoting enclosed stellar mass contained in stars less massive than the average mass (i.e., $M_{*}<0.4\, \rm M_{\odot}$), show that the low-mass stars are definitely much more spread out  in case of primordially mass-segregated clusters when compared to the unsegregated cluster.}

\end{figure}

\begin{figure}[ht]
\begin{center}
\plotone{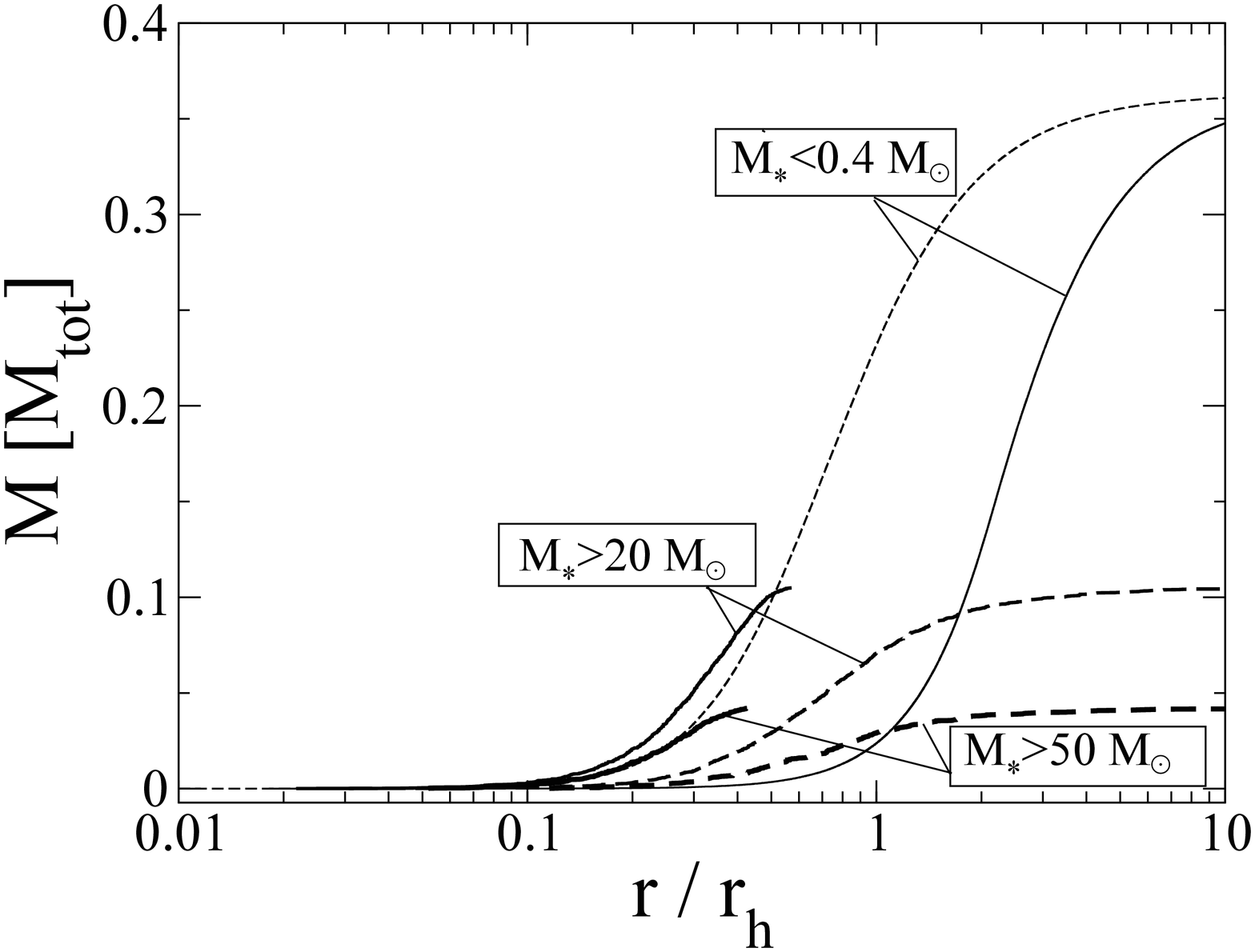}
\end{center}

\caption{\label{fig:Enclosed-mass1}Same as Fig.\ref{fig:Enclosed-mass2} but for primordially mass-segregated clusters using \citet{Baum08} recipe. The black solid lines denote the primordially
mass-segregated clusters and the dashed lines are for unsegregated clusters. Massive stars
in unsegregated clusters are at significantly larger radii, with their
median position differing by a factor of $\sim4$ for stars with $M_{*}>50\, \rm M_{\odot}$,
and by a factor of $\sim3$ for stars with $M_{*}>20\, \rm M_{\odot}$, compared to the primordially
mass-segregated clusters.}

\end{figure}

Figures \ref{fig:Average-mass-profile2} and \ref{fig:Average-mass-profile1}
show the average mass profile for the initially mass-segregated clusters (and unsegregated clusters also for comparison) using the \citet{Subr08} and \citet{Baum08} recipes, respectively. As can be seen in these plots, the average mass rises rapidly within one half-mass radius, by up to a factor 
of $\sim5$
(for the Baumgardt et al. (2008) recipe as well as 
for $S=0.3$) for the initially mass-segregated clusters.
Similarly Figures \ref{fig:Enclosed-mass2} and \ref{fig:Enclosed-mass1} show the enclosed mass for the same two prescriptions. Comparing the median 
positions of the massive stars in these models, it becomes clear that the massive stars are on average at much smaller radii in 
initially mass-segregated clusters compared to non-segregated ones, as expected.
The median distance from the cluster center can be up to a factor $\simeq$ 4 smaller for stars with mass larger than $50\,\rm
M_{\odot}$.
Comparing the enclosed stellar mass profiles for the two recipes (Figure \ref{fig:Enclosed-mass2} and Figure \ref{fig:Enclosed-mass1}) it can be seen that,  
the  \citet{Baum08} recipe agree pretty closely with the \citet{Subr08} recipe for $S=0.3$.  

 In this paper we also study the effect of flatter IMFs on the runaway collision scenario in young massive star clusters. 
We introduce a new IMF in this paper motivated by the Arches IMF from \citet{dib2007}. For the rest of the paper, this new IMF will called the ``Variable IMF" and we will
denote the sections of this IMF ($dN/dm \propto m^{-\alpha}$) as $[\alpha_1 \ldots \alpha_4]$ with $\alpha_4$
corresponding to the high mass tail of the IMF. This Variable IMF is a variation on the Arches IMF \citep{dib2007} by leaving out the middle section from $3.0 -
15.0\,\rm{M_{\odot}}$ so that the IMF is then more easily compared to the standard
Kroupa IMF. The upper mass section of the Variable IMF is much flatter than
traditional \citet{Sal55}, \citet{kroupa02}, or \citet{MilSc79} IMFs.  We 
therefore expect a greater number of high mass stars, increasing both the
average mass as well as the entire cluster mass.

\subsection{Initial Conditions}

For all the simulations in this paper our code includes physical processes such as two-body relaxation and physical stellar collisions between stars in the `sticky sphere' approximation.
In this paper we do not include stellar evolution in our calculations since our aim here to investigate dynamical processes taking place before even the most massive main sequence stars in the cluster have evolved (i.e., $< 3\,\rm Myr$). We have also not included any primordial binaries in our simulations just to keep the simple picture of runaway in presence of primordial mass segregation, and exclude the effects of primordial binaries on runaway collisions \citep{GU06}.

The most important physical properties of all our initial cluster models are given in Table \ref{t2} and Table \ref{t3}.  A comparison of the variable IMF with the Arches IMF or with standard Kroupa IMF is given in Table \ref{IMF}. In Table \ref{t2} we have listed simulations of clusters with the Variable IMF, varying $\alpha_3$ and $\alpha_4$. We have also listed the simulations done with Salpeter IMF varying $M_{\rm max}$. In all these simulations our focus has been to investigate the effect of IMF on core-collapse of young massive clusters. All our models start with an isolated Plummer sphere and the virial radius has been chosen such that the corresponding cluster has a core-collapse time of $3\,\rm Myr$. 


 In Table \ref{t3} we have listed simulations of clusters for different values of initial N and initial virial radius, $r_{\rm vir}$, varying the degree of primordial mass-segregation. 
 Virial radius defined by 
\begin{equation}
r_{\rm vir}=-\frac{GM^2}{4E}
\end{equation}
where $M$ is the total mass and $E$ is the total gravitational energy of the cluster.
 This grid formed by the different values of initial N and initial $r_{\rm vir}$ is referred to as the parameter space in the later part of the paper. For all the simulations listed here our aim has been to investigate the effect of primordial mass-segregation on $t_{\rm cc}$ of young clusters with different initial conditions ($N$ and $r_{\rm vir}$).
 We have calculated $t_{\rm cc}$ based on the central relaxation time $t_{\rm rc}$ defined as
\begin{equation}
t_{\rm rc}(0)=\frac{\sigma_{c}^3}{4.88\pi G^2 (\ln \gamma_{c} N)n \left<m\right>_{c}^2}
\end{equation}
where $\sigma_c$, n, $ \left<m\right>_{c}$  are the three-dimensional \mbox{velocity} dispersion, number density, and average stellar mass respectively at the cluster center \citep[eq. 3.37]{Spi1987}.
As a typical reference model similar to Paper I, all the simulations in Table \ref{t3} are initiated with Plummer models and Salpeter IMF (within $\rm M_{\rm min}=0.2\, \rm M_{\rm \odot}$ and $\rm M_{\rm max}=120\, \rm M_{\rm \odot}$). Another reason for
 using Plummer sphere is that the \citet{Subr08} formalism starts with an isolated Plummer sphere.

\section{Results}
\subsection{Primordial Mass segregation}


 \citet{freitag06} were the first to study the core collapse and collisional runaway for unsegregated clusters using a MC simulation code including stellar collisions. Similar to \citet{freitag06} we did not take into account stellar evolution in the simulations since all the simulations were limited to the first $3\, \rm Myr$, before the most massive stars lose mass in supernovae explosions. If a cluster had a core collapse time more than $3\, \rm Myr$ we implicitly took stellar evolution into account by ending the simulation at $3\, \rm Myr$. Since our code now includes stellar collisions, we first checked whether we are able to reproduce their results. 
 They found that for all models with $t_{\rm rc}< 20\, \rm Myr$, runaway formation of a VMS occured. 

\begin{figure}[ht]
\begin{center}
\plotone{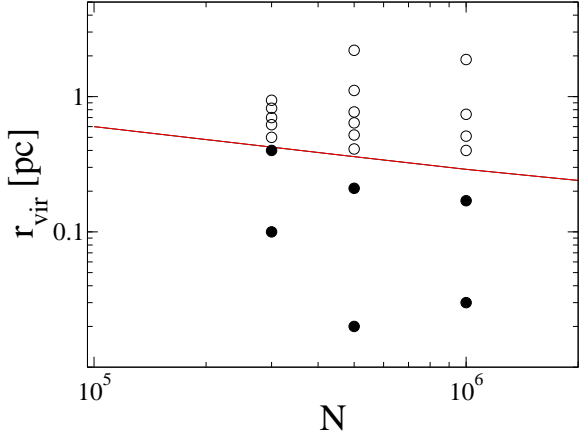}
\caption{\label{fig:parameterspace-unseg}Parameter space survey for the primordially unsegregated clusters. All the simulations shown here contain $5\times10^5$ stars with a Salpeter IMF (within $\rm M_{\rm min}=0.2\, \rm M_{\rm \odot}$ and $\rm M_{\rm max}=120\, \rm M_{\rm \odot}$). Each circle represents a simulation of a cluster with a particular virial radius ($r_{\rm vir}$) and number of stars (N). Filled circles represent simulations that result in a runaway while the empty circles indicate cases in which no runaway occur. The straight line denotes $t_{\rm cc}=0.15 t_{\rm rc}=3\,\rm Myr$ as a function of $r_{\rm vir}$ and $N$; clusters with initial conditions below this line have $t_{\rm cc} < 3\, \rm Myr$. As expected, we see in this plot that all the simulations that result in a runaway do in fact fall below this line.  
}
\end{center}
\end{figure}

 Figure \ref{fig:parameterspace-unseg} is a parameter survey similar to \citet[][their Fig. 1]{freitag06}, showing for each simulation whether a runaway occurred (filled circles) or not (open circles), for all our unsegregated models varying $r_{\rm vir}$ and the number of stars in the cluster. The solid straight line corresponds to $t_{\rm cc}=0.15t_{rc}=3\,\rm Myr$, not including collisions. Simulations of clusters with initial conditions lying below the straight line will have  $t_{\rm cc} < 3\, \rm Myr$, whereas $t_{\rm cc} > 3\, \rm Myr$ for simulations with initial conditions above the line. It can be clearly seen in Figure \ref{fig:parameterspace-unseg} that the initial conditions leading to a runaway fall indeed below this line. Thus we have successfully reproduced the results of \citet{freitag06} with our code and have reconfirmed the validity of the simple criterion for runaway collisions to occur in young dense star clusters.

\begin{figure}[ht]
\begin{center}
\plotone{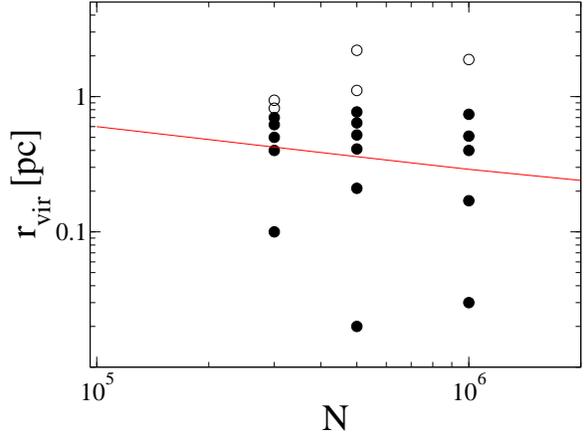}
\caption{\label{fig:parameterspace-seg}Same as Fig.\ref{fig:parameterspace-unseg} but all the simulations are for primordially mass-segregated clusters ($S=0.3$). The solid straight line once again denotes denotes $t_{\rm cc}=0.15 t_{\rm rc}=3\,\rm Myr$ as a function of $r_{\rm vir}$ and $N$ for initially unsegregated clusters. This plot show that simulations of primordially mass-segregated clusters with initial conditions even above the solid straight line, might have core collapse time less than $3\, \rm Myr$ and can result in a runaway. Thus, simulations of primordially mass-segregated clusters increase the parameter space (range of $r_{\rm vir}$ in pc) for runaway collisions to happen when compared to simulations of primordially unsegregated clusters by at least a factor of $\approx 3$. }
\end{center}
\end{figure}

 We then repeated the same set of simulations for primordially mass-segregated clusters using the recipe from \citet{Subr08} with $S > 0$. 
Figure \ref{fig:parameterspace-seg} shows our results for mass-segregation parameter $S=0.30$. We see that, with primordial mass segregation, the initial cluster $r_{\rm vir}$ can be chosen several times larger compared to unsegregated clusters and still lead to a runaway.
Thus, simulations of primordially mass-segregated clusters increase the parameter space (range of $r_{\rm vir}$ in pc) for runaway collisions to happen when compared to simulations of primordially unsegregated clusters.

\begin{figure}[ht]
\plotone{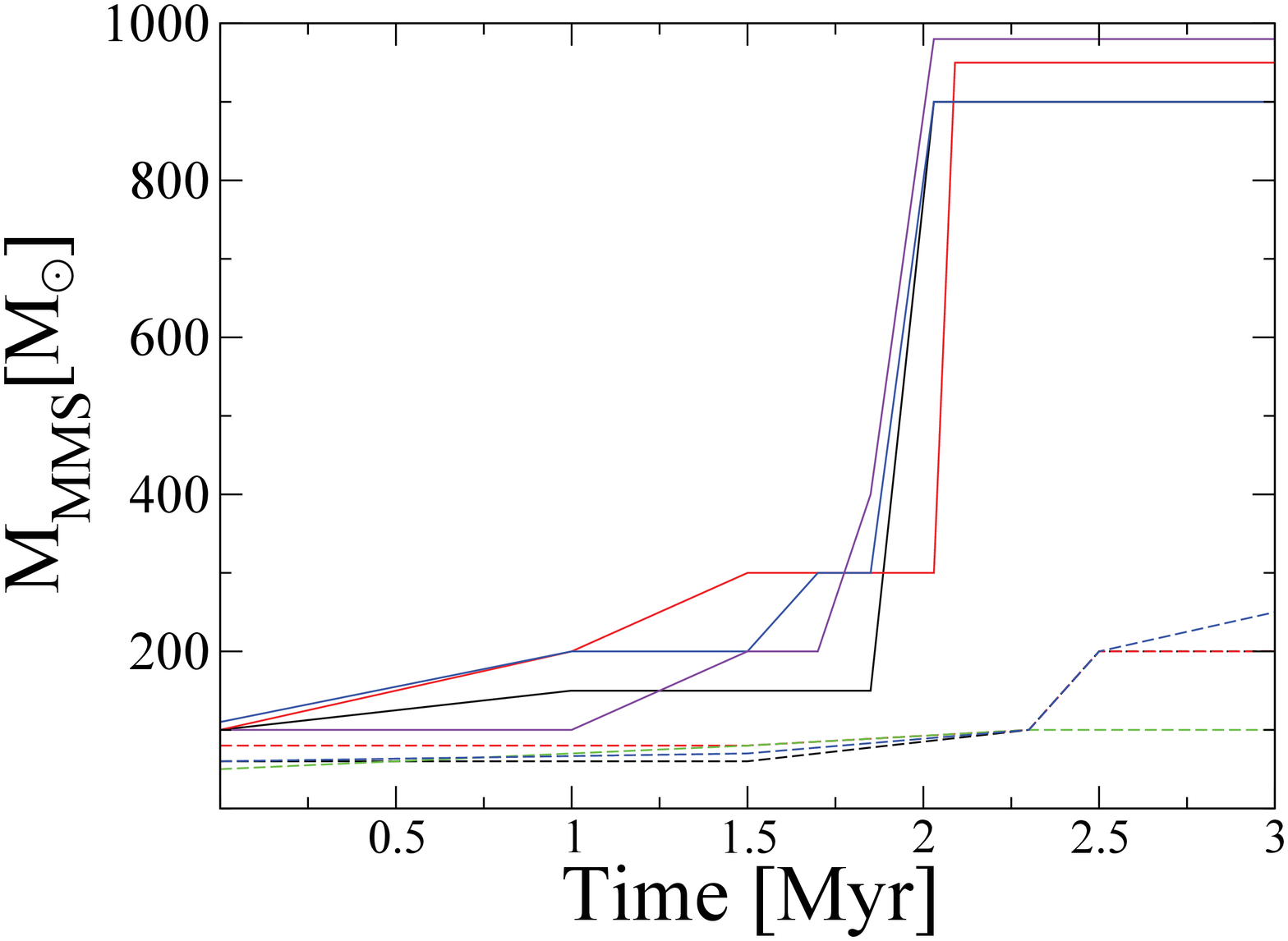}
\caption{\label{fig:Mergertree2}Growth curves of the most massive star ($\rm M_{\rm MMS}$) of an unsegregated cluster (dashed lines) and primordially mass-segregated cluster (solid lines) with $N=5\times10^{5}$ stars and virial radii of $0.64\, \rm pc $. The different colors indicate simulations of the same cluster with different random seeds. While the most massive star in the primordially mass-segregated cluster have a very steep and rapid mass growth leading to a runaway, the most massive star in the unsegregated cluster shows no such effect. For the unsegregated cluster $t_{\rm rc}=38.67$(red), $38.33$(blue), $38.66$(green) and $40\,\rm Myr$(purple) while for the primordially mass-segregated cluster $t_{\rm rc}$ values are $13.53$(red), $14.0$(blue), $14.0$(green) and $13.33\,\rm Myr$(purple).}
\end{figure}

To further illustrate the effect primordial mass segregation has on the mass growth of the most massive star we compared results of a primordially mass-segregated vs an unsegregated cluster, with otherwise identical initial conditions. Figure \ref{fig:Mergertree2} shows the growth curve of the most massive star in an unsegregated  cluster and in a primordially mass-segregated cluster ($S=0.3$), with $N=5\times 10^{5}$ stars and $r_{\rm vir}=0.64\, \rm pc$. 
The unsegregated cluster does not have a runaway or a very steep mass growth before it reaches core collapse, and only a $200\, \rm M_{\odot}$ star is formed within $3\, \rm Myr$.  For the primordially mass-segregated cluster, we clearly see a very steep mass growth leading to a formation of a $900\, \rm M_{\odot}$ star within $3\,\rm Myr$ and hence a runaway. A general result, in agreement with \citet{freitag06}, is that only one VMS forms in the cluster and there is no sign of multiple runaways. Note that this result might be different for clusters with significant fractions of primordial binaries \citep{GU06}. As in this example, we also note in all our simulations that the time of the runaway coincides with the time of core-collapse, which decreases from $t_{cc}\sim 6\,\rm Myr$ for the unsegregated cluster, and to $t_{\rm cc}\sim2\,\rm Myr$ for the primordially mass-segregated cluster.

\begin{figure}[ht]
\begin{center}
\plotone{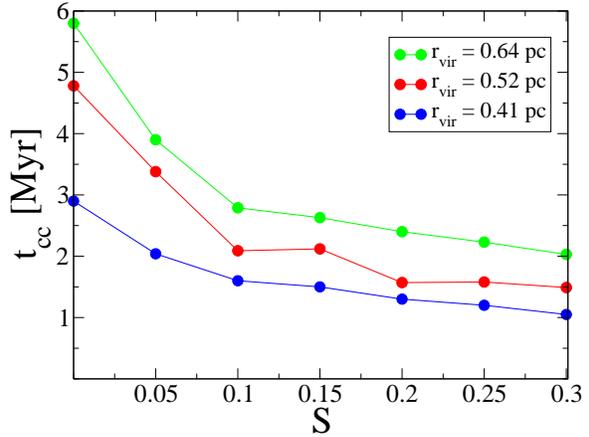}
\end{center}

\caption{\label{fig:corecollapse}Decrease in the core collapse time ($t_{\rm cc}$) of a cluster with $N=5 \times 10^{5} $ stars, as the degree of primordial mass segregation ($S$) is increased in the cluster. For each value of $S$, the three different colors represent simulations with three different virial radii ($r_{\rm vir}$).}

\end{figure}

To show how strongly the core collapse time decreases for other values of $S$ we plot 
 in Figure \ref{fig:corecollapse}  the core collapse time for clusters with different initial $r_{\rm vir}$, against $S$. We clearly see a trend of decreasing $t_{\rm cc}$ with increasing $S$. The decrease in $t_{\rm cc}$ is steepest for $S\lesssim 0.1$, causing a reduction of $t_{\rm cc}$ by a factor of $\approx 2$. For $S>0.1$ the decrease in $t_{\rm cc}$ is weaker, reducing $t_{\rm cc}$ by another $30\%$. This trend is also illustrated in  Table \ref{t3} where we have shown that $t_{\rm cc}$ decreases for simulations with primordial mass-segregation, when compared to simulations of initially unsegregated clusters with similar initial conditions.

\begin{figure}[ht]
\begin{center}
\plotone{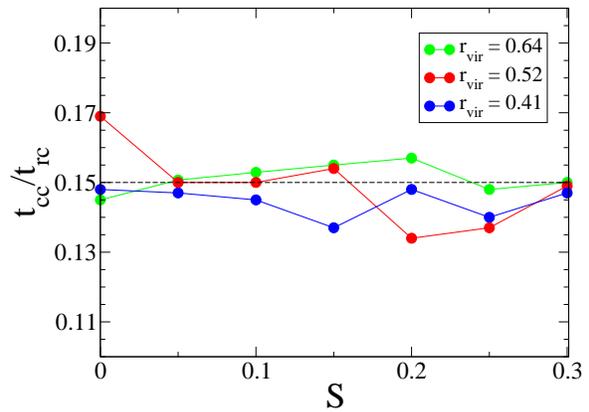}
\caption{\label{fig:ratio}Dependence of core collapse time in units of the central relaxation time, on primordial mass segregation ($S$). For each value of $S$, the three different colors represent simulations with three different virial radii ($r_{\rm vir}$). For comparison, $t_{\rm cc}/t_{\rm rc}$ as predicted in Paper I for unsegregated clusters is plotted as the dotted line. Even for the mass segregated clusters $t_{\rm cc}/t_{\rm rc}$ remains  roughly consistent with Paper I.}
\end{center}
\end{figure}

\begin{figure}[ht]
\begin{center}
\plotone{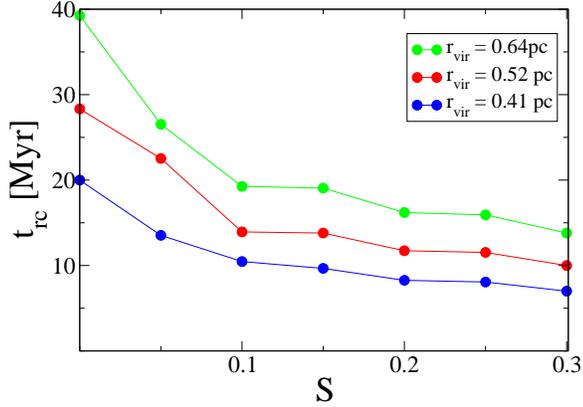}
\caption{\label{fig:trc}Decrease in the central relaxation time ($t_{\rm rc}$ in $\rm Myr$) of a cluster of $5 \times 10^{5} $ stars with the increase in the degree of primordial mass segregation ($S$). For each value of $S$, the three different colors represent simulations with three different virial radii ($r_{\rm vir}$).}
\end{center}
\end{figure}

 To analyze this trend further we first check whether the simple relation from Paper I between $t_{\rm cc}$ and $t_{\rm rc}(0)$ ($t_{\rm cc}/t_{\rm rc}(0)=0.15$) remains still valid for mass segregated clusters. In Figure \ref{fig:ratio} we plot $t_{\rm cc}/t_{\rm rc}$ against S for clusters with different $r_{\rm vir}$. We see that $t_{\rm cc}/t_{\rm rc}$ remains nearly constant at $0.15$ with only a $\sim10\%$ scatter. Thus, we conclude, that the ratio of $t_{\rm cc}/t_{\rm rc}$ remains consistent with the value found in Paper I and the strongly decreasing $t_{\rm cc}$ can then only be caused by a decrease in $t_{\rm rc}(0)$. This is also shown in Figure \ref{fig:trc}, where we plot $t_{\rm rc}(0)$ against $S$ for clusters with different $r_{\rm vir}$. We notice a very similar decrease in $t_{\rm rc}$ as in $t_{\rm cc}$, with increasing $S$, so the reason for the shorter $t_{\rm cc}$ must be attributed to shorter $t_{\rm rc}$.  

\begin{figure}[ht]
\begin{center}
\plotone{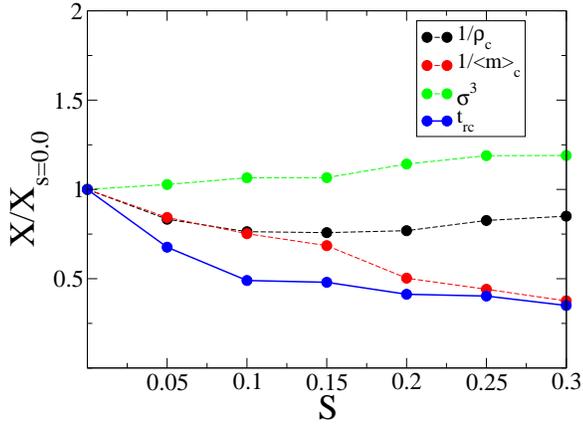}
\caption{\label{fig:rhoavgm}Evolution of the central relaxation time $t_{\rm rc}$ (solid line) and the factors affecting $t_{\rm rc}$ (dotted lines). All parameters are normalized to their values for an unsegregated cluster, ($\rm X/X_{S=0.0}$). The black line is for the inverse of central density ($1/\rho_{c}$), the intermediate grey is for the inverse of average stellar mass in the core ($1/\left<m\right>_{c}$), and the light grey represents the cube of the velocity dispersion in the core ($\rm \sigma^3_{c}$).  While $\rm \sigma^3_{c}$ remains almost the same with increasing $S$, both $1/\rho_{c}$ and $1/\left<m\right>_{c}$ decrease $t_{\rm rc}$. However, for higher values of $S$ ($S>0.15$), it is mainly $1/\left<m\right>_{c}$ which decreases $t_{\rm rc}$.}
\end{center}
\end{figure}

However, in the \citet{Subr08} formulation it is not a priori clear whether this decrease in $t_{\rm rc}$ is entirely related to an increase in primordial mass segregation, as the central density ($\rho_{\rm c}$) also increases with $S$ ($\S4.2$). In order to disentangle these effects, we calculate relative contributions of all the factors to $t_{\rm rc}$. We note that $t_{\rm rc} \propto \sigma_c^3/ \rho_{c}\left<m\right>_c$ where $\sigma_{c}$ is the velocity dispersion in the core, and $\left<m\right>_{c}$ is the average stellar mass in the core.
Figure \ref{fig:rhoavgm} plots the variation of $\sigma_{c}^3$, $1/\rho_{c}$ and $1/\left<m\right>_c$, normalized to their values at $S=0$, against $S$ in a cluster with $N=5\times10^5$ stars. The inverse of the normalized initial central density of the clusters decreases as $S$ is increased from $0.0$ to $0.15$ by $25\%$, and then remains nearly constant for larger S values. The inverse of normalized $\left<m\right>_c$ in the cluster always decreases with the increase in mass segregation, up to a factor of $ 2.5$ for $S=0.3$. The variation of $\sigma_{c}^3$, on the other hand, is only by a factor of $\approx 10\%$, implying that $\sigma_{c}$ stays almost constant in the core and does not change with $S$. From this it follows that for $S<0.15$ both the increase in $\left<m\right>_{c}$ as well as $\rho_{c}$ contributes equally to the decrease in $t_{\rm rc}$. On the other hand, for larger $S$, $t_{\rm rc}$ is dominated by $\left<m\right>_{c}$ while the contribution from $\rho_{c}$ is minimal ($\approx 20\%$). So it is the increase in $\left<m\right>_{c}$ that mainly causes the low $t_{\rm rc}$ values for larger $S$. Thus the maximum increase in the parameter space for a runaway to occur is driven mainly by ``true" primordial mass segregation.

\begin{figure}[ht]
\begin{center}
\plotone{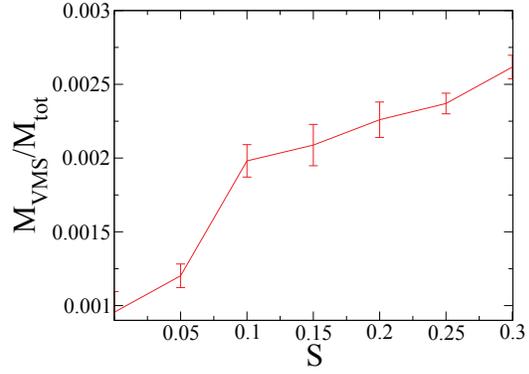}
\caption{\label{fig:svsmass}Fraction of the total stellar mass in the cluster, that ends up as a VMS after a runaway collision, as a function of the degree of mass segregation. All the simulations are for a cluster of $5\times10^5$ stars and a virial radius of $0.64\, \rm pc $. The $M_{\rm VMS}/M_{\rm tot}$ shown here is averaged over 5 simulations with the same initial conditions and different random seeds. The error bars are the standard deviations calculated from the data and the average value. }
\end{center}
\end{figure}

The advantage of including stellar collisions in our code is that we are able to directly determine how much mass eventually ends up in the VMS. 
 Paper I showed that, as the core collapse proceeds, the mass contained in the collapsing core which forms the mass reservoir for the runaway converges to a value $M_{\rm cc}\sim 0.001-0.002$ $M_{\rm tot}$, where $M_{\rm tot}$ denotes the total mass of the cluster. In general we find that the fraction of $ M_{\rm tot}$ that ends up as a VMS is in a similar mass range as $M_{\rm cc}$ in Paper I. In Figure \ref{fig:svsmass} we show that fraction as a function of $S$ for a cluster with $N=5 \times 10^5$ stars. It can be clearly seen that the fraction of $M_{\rm tot}$ ending up as a VMS increases almost by a factor of $3$ for $S=0.3$ and hence mass of the VMS increases significantly with primordial mass segregation. Thus primordial mass segregation not only increases the parameter space for runaways to occur, but also produces more massive VMS.





\subsection{Top-Heavy IMFs}
\begin{figure}[ht]
  \begin{center}
\plotone{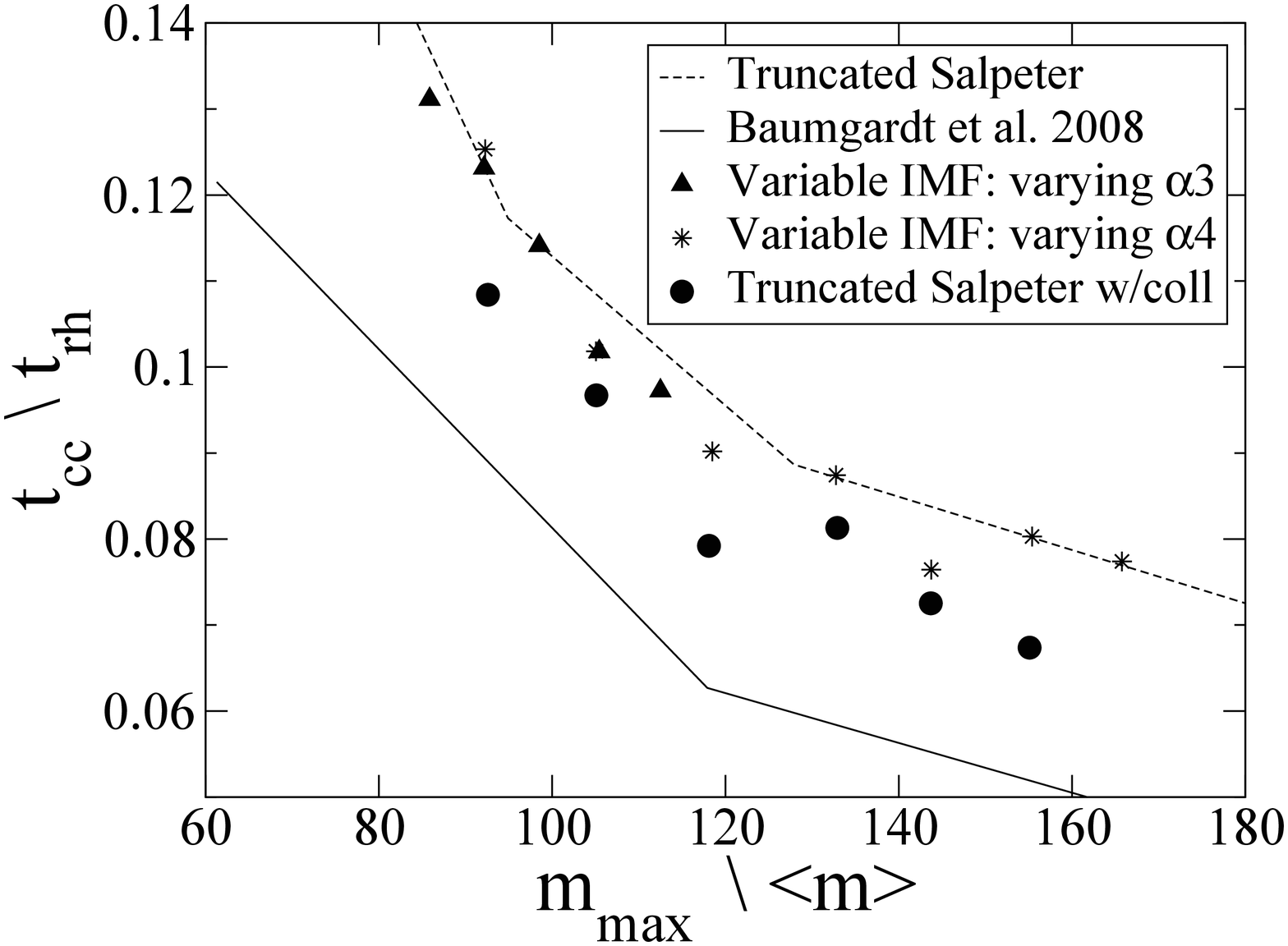}
     \caption{\label{fig:IMF-tcc}Dependence of the core collapse time in units of the half-mass relaxation time, on the shape and width of the IMF. The horizontal axis shows the ratio of maximum to mean stellar mass ($m_{max}/\left<m\right>$) in the IMF. The solid black line is for mass-segregated clusters (\citet{Baum08}), whereas all other symbols are for unsegregated clusters. The black dashed straight line, the triangles and the stars denote the simulations with physical stellar collisions turned off whereas the circles represent simulations with collisions (truncated Salpeter IMF with $M_{\rm max}=110$, $100$, $90$, $80$, $70$, $60$). The stars correspond to the Variable IMF varied as a power law from $-1.72$ to $-2.3$ ($\alpha4$) and the triangles correspond to the middle section of the Variable IMF varied as a power law from $-1.9$ to $-2.3$ ($\alpha3$). Flatter IMFs (triangles and stars) increase $t_{\rm cc}$ by $30\%$ but when collisions (circles) are included $t_{\rm cc}$ decrease on an average by $15\%$.}
  \end{center}
\end{figure}

As discussed previously, there are indications that some young massive star clusters may be born with IMFs flatter than the Salpeter IMF. Here we study how such flatter IMFs influence the available parameter space of cluster initial conditions for a runaway to occur. One crucial result from Paper I is that the ratio of maximum to average stellar mass in the IMF 
is the most important parameter setting the timescale for the onset of core collapse. From Paper I if $m_{\rm max}/\left<m\right><40$, $t_{\rm cc}\propto (m_{\rm max}/\left<m\right>)^{-1.3}$ and if $m_{\rm max}/\left<m\right>\simeq50$, a domain is reached by any realistic IMF where  $t_{\rm cc}\simeq0.15t_{\rm rc}$. A flatter IMF increases the number of massive stars, thus increasing the average stellar mass in the cluster, which in turn causes an increase in $t_{\rm cc}$. This means that, although for flatter IMFs we have more massive stars, the parameter space for runaway collisions to occur actually gets \emph{smaller}. On the other hand, collisions can reduce $t_{\rm cc}$ by constantly dissipating orbital energy  \citep{freitag06}. This means that the increase in $t_{\rm cc}$ seen in simulations without an explicit treatment of stellar collisions,  is to some degree compensated by the effect of collisions. In Figure \ref{fig:IMF-tcc} we quantify this effect for flatter IMFs considering a wide range of $m_{\rm max}/\left<m\right>$ values. We plot  $t_{\rm cc}/t_{\rm rh}$ vs $m_{\rm max}/\left<m\right>$ with and without collisions. As can be clearly seen, with stellar collisions turned off, $t_{\rm cc}$ increases; but when collisions are included, $t_{\rm cc}$ decreases on average by $15\%$. Overall this causes $t_{\rm cc}/t_{\rm rh}$ for a collisional cluster causes the core collapse to stay close to the value for the standard Salpeter IMF (0.07) all the way to $m_{\rm max}/\left<m\right> \approx140$.

We also, studied the combined effects of initial mass segregation and a flat IMF on the evolution of young massive clusters. This is mainly motivated by \citet{ES09} where the global IMF of the Arches cluster is determined to be flatter than Salpeter, while Arches is also known to be strongly mass segregated. Note however that the uncertainity on the slope is such that it may not be incompatible with a Salpeter IMF and there are models of Arches \citep{Harfst10} that can explain the degree of mass segregation with a Salpeter IMF and unsegregated initial conditions. On the other hand \citet{Harfst10} do not rule out the possibility that Arches could be primordially mass-segregated. In Figure \ref{fig:IMF-tcc}, we show the results for primordially mass-segregated clusters using the \citet{Baum08} recipe, which, unlike the \citet{Subr08} recipe, can be applied to clusters with arbitrary density profiles. This way the decrease in $t_{\rm cc}$ is only caused by the increase in the average stellar mass at the center of the cluster. From Figure \ref{fig:IMF-tcc} we see that, as in the unsegregated case, the effect of a flatter IMF increases $t_{\rm cc}$, and for $m_{\rm max}/\left<m\right>=120$ is very close to the standard Salpeter value. The change for $m_{\rm max}/\left<m\right>=160$ is only by $20\%$. Note that $m_{\rm max}/\left<m\right>=120$ corresponds to  either a Salpeter IMF with a high mass cut-off at $80\, \rm M_{\odot}$, or to an IMF with a slope of $-2.1$ and a full mass spectrum (from $0.1- 120\, \rm M_{\odot}$). This latter slope is precisely the one suggested for the global IMF of Arches by \citet{ES09}. If the global IMF of the Arches star cluster is also representative of the IMFs for much more massive clusters like the Westerlund 1, this would imply that the simple relation from Paper I can be used to determine whether a young massive cluster will undergo a runaway or not, without explicitly accounting for a different global IMF or primordial mass segregation.

\section{Summary and Discussions}

This work is a continuation of our study of the runaway collision scenario in young dense star clusters. In this paper our goal was to investigate the parameter space for runaway collisions to occur, when the effects of stellar collisions, primordial mass segregation and a globally flatter IMF (as indicated by observations of young massive star clusters) are accounted for. We considered clusters having Plummer density profiles, varying the initial virial radius, the number of stars in the cluster and the IMF. Primordial mass segregation was generated using the \citet{Subr08} and \cite{Baum08} methods. We naturally expected that primordial mass segregation in a cluster should lead to shorter core-collapse times since the massive stars start their lives closer to the center of the cluster, where the density and therefore collision rates are highest. Indeed, from our simulations we found that the core-collapse time, $t_{\rm cc}$ decreases with increasing the degree of primordial mass segregation. increasing the initial virial radius for runaway collisions to happen, by at least a factor of $\approx 3$.

We find that the simple relation between core-collapse and central relaxation times, ($t_{\rm cc}=0.15t_{\rm rc}$), discussed in Paper I, still holds even with primordial mass segregation. The strong decrease in $t_{cc}$ hence implies a similar reduction in $t_{\rm rc}$, which accelerates the core collapse and enlarges the parameter space for runaway collisions to occur. This decrease in $t_{\rm rc}$, in both the \citet{Subr08} and \citet {Baum08} prescriptions, is caused by the strong increase in the average stellar mass in the core. In the \citet{Subr08} recipe the decrease in $t_{\rm rc}$ is also caused by an increase in central density, which, however, does not contribute much at higher degrees of mass segregation ($S>0.15$).

 We find that the fraction of the total cluster mass that is eventually accumulated onto the $\rm VMS$, the possible progenitor of an IMBH is comparable to the mass of the collapsing core, $M_{\rm cc}$, already determined in Paper I. If IMBHs were indeed formed in massive star clusters, our results show that the ratio of the mass of the IMBH to the total stellar cluster mass follow the $M/M_{\rm bulge}$ ratio.
  However, the runaway collision scenario imposes the requirement that the IMBH must have formed in a cluster with initial central relaxation time shorter than $20\, \rm Myr$. 

 Finally we find that for top-heavy IMFs (in unsegregated clusters) the parameter space for runaway collisions is \emph{reduced}, since flatter IMFs increase $t_{\rm cc}$, by a factor $\approx 1.4$ for $m_{\rm max}/\left<m\right>\approx 120$. However this increase is to some degree balanced by collisions, which happen more frequently in clusters with flatter IMFs as we get more massive stars. In addition, primordial mass segregation in these clusters again reduces $t_{\rm cc}$, and for an IMF with a slope of $-2.1$, as may be present in the Arches cluster \citep{ES09}, this reduced $t_{\rm cc}$ is nearly the same as for an unsegregated cluster with a Salpeter IMF. Thus, if the IMF is Salpeter like, primordial mass-segregation increases the parameter space for runaway collisions to occur, whereas for flatter IMFs the parameter space remains very similar to that in unsegregated clusters with a Salpeter like IMF. From this we can conclude that if young massive star clusters (like Westerlund 1) are primordially mass segregated and have an IMF slope similar to Arches then the results from Paper I are directly applicable.


We thank  John Fregeau and Sourav Chatterjee for the useful discussions.  This work was supported by NSF Grant AST-060.7498 and NASA Grant NNX08AG66G at Northwestern University. 

\clearpage



\bibliographystyle{apj}
\bibliography{runaway}

\begin{deluxetable}{crrr}  
\tablecolumns{3}
\tablewidth{0pc}
\tablecaption{CLASSIFICATION OF IMF \label {IMF}}
\tablehead{   
 \colhead{IMF} &
 \colhead{Range of m}  & 
 \colhead{$\alpha$} \\
 \colhead{   } &
 \colhead{($M_{\odot}$)}  & 
 \colhead{    }

 }
\startdata
Kroupa    &  &  & \\  
                 & $0.1-0.5$   & -1.3\\
  & $0.5-\infty$   & -2.3\\
  \hline
Arches    &  &  & \\  
 & $0.1-0.5$         & -1.3\\
 & $0.5-1.0$         & -2.3\\
 & $1.0-3.0$         & -2.04\\
    & $3.0-15.0$      & -1.5\\
 & $15.0-\infty$   & -1.72\\
\hline
Variable    &  &  & \\  
 & $0.1-0.5$   & -1.3\\
 & $0.5-1.0$   & -2.3\\
  & $1.0-3.0$   & -2.04\\
  & $3.0-\infty$   & -1.72\\

\enddata
\end{deluxetable}

\begin{deluxetable}{lllcccc}
\tablecolumns{4}
\tablewidth{0pc}
\tablecaption{INITIAL CONDITIONS AND RESULTS FOR SIMULATIONS WITH VARYING IMF.\label{t2}}
\tablehead{
\colhead{Name} & \colhead{IMF} & \colhead{$r_{\rm vir}$} & \colhead{$t_{cc}/t_{rh}$} \\
\colhead{} & \colhead{} & \colhead{$(pc)$} & \colhead{} \\
}
\startdata


01 & Variable $\alpha_3=-2.04, \alpha_4=-1.7$ & 0.39 & 0.127 \\
02 & Variable $\alpha_3=-2.04, \alpha_4=-1.8$ & 0.39 & 0.103 \\
03 & Variable $\alpha_3=-2.04, \alpha_4=-1.9$ & 0.39 & 0.092 \\
04 & Variable $\alpha_3=-2.04, \alpha_4=-2.0$ & 0.39 & 0.088 \\
05 & Variable $\alpha_3=-2.04, \alpha_4=-2.1$ & 0.39 & 0.077 \\
06 & Variable $\alpha_3=-2.04, \alpha_4=-2.2$ & 0.39 & 0.081 \\
07 & Variable $\alpha_3=-2.04, \alpha_4=-2.3$ & 0.39 & 0.078 \\
08 & Variable $\alpha_3=-1.9, \alpha_4=-1.72$ & 0.39 & 0.132 \\
09 & Variable $\alpha_3=-2.0, \alpha_4=-1.72$ & 0.39 & 0.125 \\
10 & Variable $\alpha_3=-2.1, \alpha_4=-1.72$ & 0.39 & 0.117 \\
11 & Variable $\alpha_3=-2.2, \alpha_4=-1.72$ & 0.39 & 0.102 \\
12 & Variable $\alpha_3=-2.3, \alpha_4=-1.72$ & 0.39 & 0.099 \\
13 & Salpeter $M_{\rm max}=110$ & 0.39 & 0.067 \\
14 & Salpeter $M_{\rm max}=100$ & 0.39 & 0.073 \\
15 & Salpeter $M_{\rm max}=90$ & 0.39 & 0.0812\\
16 & Salpeter $M_{\rm max}=80$ & 0.39 & 0.079 \\
17 & Salpeter $M_{\rm max}=70$ & 0.39 & 0.097 \\
18 & Salpeter $M_{\rm max}=60$ & 0.39 & 0.108\\
\enddata

\tablenotetext{a}{All models start as Plummer spheres with $N=5\times 10^{5}$ and the Variable IMF covers the same mass range ($0.2-120\rm M_{\odot}$).} 
\tablecomments{$r_{\rm vir}$ is the virial radius of the cluster, $t_{\rm cc}$ denotes the core collapse time of the cluster and $t_{\rm rh}$ denotes the half mass relaxation time.}

\end{deluxetable}

\begin{deluxetable}{ccccccc}  
  \tablecolumns{6} 
  \tablewidth{0pc}
  \tablecaption{INITIAL CONDITIONS AND RESULTS OF SIMULATIONS WITH PRIMORDIAL MASS SEGREGATION ($\breve{S}$UBR RECIPE).\label{t3}}

\tablehead{   
 \colhead{Name}&
  \colhead{$N$} &
  \colhead{$r_{\rm vir }$} &
  \colhead{$S$} &
  \colhead{$t_{\rm rc}(0)$} &
  \colhead{$t_{\rm cc}$} \\
  \colhead{} &
  \colhead{} &
  \colhead{(pc)} &
  \colhead{} &
  \colhead{   (Myr)}&
  \colhead{(Myr)}
}
\startdata

1a & 3x$10^{5}$ & 0.50 & 0.00 & 29.30 & 4.40\\
1b & 3x$10^{5}$ & 0.50 & 0.05 & 21.86 &3.28\\
1c & 3x$10^{5}$ & 0.50 & 0.10 & 19.33 & 2.90\\
1d & 3x$10^{5}$ & 0.50 & 0.20 & 11.26 & 1.69\\
1e & 3x$10^{5}$ & 0.50 & 0.30 & 6.80 & 1.02 \\
2a & 3 x$10^{5}$ & 0.62 & 0.00 & 28.66 & 4.30\\  
2b & 3x$10^{5}$ & 0.62 & 0.05 & 24.53 & 3.68\\
2c & 3x$10^{5}$ & 0.62 & 0.10 & 13.00 & 1.95\\
2d & 3x$10^{5}$ & 0.62 & 0.20 & 10.33 & 1.55\\
2e & 3x$10^{5}$ & 0.62 & 0.30 & 5.86 & 0.88\\
3a & 3 x$10^{5}$ & 0.77 & 0.00 & 41.33 &  6.20\\  
3b & 3x$10^{5}$ & 0.77 & 0.05 & 31.53 & 4.73\\
3c & 3x$10^{5}$ & 0.77 & 0.10 & 20.80 & 3.12\\
3d & 3x$10^{5}$ & 0.77 & 0.20 & 19.20 & 2.88\\
3fe& 3x$10^{5}$ & 0.77 & 0.30 & 13.93 & 2.09\\

 5a & 5x$10^{5}$ & 0.41 & 0.00 & 25.86 & 3.88\\
 5b & 5x$10^{5}$ & 0.41 & 0.05 &  21.13 & 3.17\\
 5c & 5x$10^{5}$ & 0.41 & 0.10 &  17.4 & 2.61\\
 5d & 5x$10^{5}$ & 0.41 & 0.15 & 10.00 & 1.50\\
 5e &  5x$10^{5}$ & 0.41 & 0.20 & 8.66 & 1.30\\
 5f &5x$10^{5}$ & 0.41 & 0.25 &  8.0 & 1.20\\
 5g &5x$10^{5}$ & 0.41 & 0.30 &  5.53 & 0.83\\
 6a & 5x$10^{5}$ & 0.52 & 0.00 &  31.86 & 4.78\\  
 6b & 5x$10^{5}$ & 0.52 & 0.05 &  22.53 & 3.38\\
 6c & 5x$10^{5}$ & 0.52 & 0.10 &  13.93 & 2.09\\
 6d & 5x$10^{5}$ & 0.52 & 0.15 & 14.13 & 2.12\\
 6e & 5x$10^{5}$ & 0.52 & 0.20 & 10.46 & 1.57\\
 6f & 5x$10^{5}$ & 0.52 & 0.25 & 10.53 & 1.58\\
 6g & 5x$10^{5}$ & 0.52 & 0.30 & 9.93 & 1.49\\
 
 7a & 5x$10^{5}$ & 0.64 & 0.00 &  38.66 & 5.80\\ 
 7b & 5x$10^{5}$ & 0.64 & 0.05 &  23.33 & 3.50 \\
 7c & 5x$10^{5}$ & 0.64 & 0.10 &  16.66 & 2.50\\
 7d & 5x$10^{5}$ & 0.64 & 0.15 &  17.53 & 2.63\\
 7e & 5x$10^{5}$ & 0.64 & 0.20 &  16.00 & 2.40\\
 7f & 5x$10^{5}$ & 0.64 & 0.25 &  16.00 & 2.40\\
 7g & 5x$10^{5}$ & 0.64 & 0.30 &  13.53 & 2.03\\
 7h & 5x$10^{5}$ & 0.64 & 0.00 &  38.33 & 5.75\\ 
 7i & 5x$10^{5}$ & 0.64 & 0.05 &  23.33 & 3.50 \\
 7j & 5x$10^{5}$ & 0.64 & 0.10 &  15.93 & 2.39\\
 7k & 5x$10^{5}$ & 0.64 & 0.15 &  17.86 & 2.68\\
 7l & 5x$10^{5}$ & 0.64 & 0.20 &  16.00 & 2.40\\
 7m & 5x$10^{5}$ & 0.64 & 0.25 &  15.93 & 2.39\\
 7n & 5x$10^{5}$ & 0.64 & 0.30 &  14.00 & 2.10\\
 7o & 5x$10^{5}$ & 0.64 & 0.00 &  38.66 & 5.80\\ 
 7p & 5x$10^{5}$ & 0.64 & 0.05 &  25.33 & 3.80 \\
 7q & 5x$10^{5}$ & 0.64 & 0.10 &  16.66 & 2.50\\
 7r & 5x$10^{5}$ & 0.64 & 0.15 &  10.76 & 2.69\\
 7s & 5x$10^{5}$ & 0.64 & 0.20 &  16.00 & 2.40\\
 7t & 5x$10^{5}$ & 0.64 & 0.25 &  16.00 & 2.40\\
 7u & 5x$10^{5}$ & 0.64 & 0.30 &  14.06 & 2.11\\
 7v & 5x$10^{5}$ & 0.64 & 0.00 &  40.00 & 6.00\\ 
 7x & 5x$10^{5}$ & 0.64 & 0.05 &  23.33 & 3.50 \\
 7w & 5x$10^{5}$ & 0.64 & 0.10 &  16.86 & 2.53\\
 7y & 5x$10^{5}$ & 0.64 & 0.15 &  17.53 & 2.63\\
 7z1 & 5x$10^{5}$ & 0.64 & 0.20 &  16.00 & 2.40\\
 7z2 & 5x$10^{5}$ & 0.64 & 0.25 &  16.00 & 2.40\\
 7z3 & 5x$10^{5}$ & 0.64 & 0.30 &  13.33 & 2.00\\

 8a & 5x$10^{5}$ & 0.77 & 0.00 &  52.60 & 7.89\\
 8b & 5x$10^{5}$ & 0.77 & 0.05 &  34.60 & 5.19\\
 8c & 5x$10^{5}$ & 0.77 & 0.10 &  28.26 & 4.24\\
 8d & 5x$10^{5}$ & 0.77 & 0.15 &  24.20 & 3.63\\
 8e & 5x$10^{5}$ & 0.77 & 0.20 &  20.40 & 3.06\\
 8f & 5x$10^{5}$ & 0.77 & 0.25 &  19.73 & 2.96\\
 8g & 5x$10^{5}$ & 0.77 & 0.30 & 16.53 & 2.48\\

9a & 1x$10^{6}$ & 0.41 & 0.10 &  20.8 & 3.12  \\
9b & 1x$10^{6}$ & 0.41 & 0.15 &  19.06 & 2.86  \\
9c & 1x$10^{6}$ & 0.41 & 0.20 &  16.53 & 2.48  \\
9d & 1x$10^{6}$ & 0.41 & 0.30 & 10.40 & 1.56  \\ 
10a & 1x$10^{6}$ & 0.51 & 0.10 &  32.60 & 4.89  \\
10b & 1x$10^{6}$ & 0.51 & 0.15 &  21.46 & 3.22 \\
10c & 1x$10^{6}$ & 0.51 & 0.20 &  17.66 & 2.65  \\
10d & 1x$10^{6}$ & 0.51 & 0.30 &  15.53 & 2.33 \\
11a & 1x$10^{6}$ & 0.83 & 0.10 &  46.53 & 6.98  \\
11b & 1x$10^{6}$ & 0.83 & 0.15 &  43.73 & 6.56  \\
11c & 1x$10^{6}$ & 0.83 & 0.20 &  27.60 & 4.24  \\
11d & 1x$10^{6}$ & 0.83 & 0.30 &  20.06 & 3.01  \\
12a & 1x$10^{6}$ & 1.10 & 0.10 &  60.53 & 9.08  \\
12b & 1x$10^{6}$ & 1.10 & 0.15 &  59.26 & 8.89  \\
12c & 1x$10^{6}$ & 1.10 & 0.20 &  52.40 & 7.86  \\
12d & 1x$10^{6}$ & 1.10 & 0.30 &  44.46 & 6.67  \\

\enddata

\tablecomments{  Here $N$ and $r_{\rm vir}$ are the initial number of stars and the initial virial radius respectively. $S$ is the degree of mass segregation (equation (\ref{mass-seg})), $t_{\rm rc}(0)$ is the initial central relaxation time  and $t_{\rm cc}$ denotes the core collapse time.}
\tablenotetext{a}{All initial models are isolated Plummer spheres with standard Salpeter IMFs with mass range ($0.2-120\rm M_{\odot}$).}
\tablenotetext{b}{Calculation of $t_{\rm rc}$ and $t_{\rm cc}$: For Plummer models $r_{\rm v}=1$ and $r_{\rm h}=0.769$ in $N$-body units  (Paper I, their Table 1). From $r_{\rm v}$ in physical units we calculate $r_{\rm h}$ in physical units. With initial N and initial $r_{\rm h}$ (in physical units) we also calculate the Initial half-mass relaxation time in physical units, given by $t_{\rm rh}=(0.138N/\ln\gamma N$)$\times(r_{\rm h}^3/GM)^{1/2}$. From $t_{\rm rh}$ we calculate $t_{\rm rc}$ ($t_{\rm rh}$ and $t_{\rm rc}$ for Plummer models is $0.093$ and $0.0437$ respectively, in Fokker-Planck units). Finally we obtain $t_{\rm cc}=0.15t_{\rm rc}$. }

\end{deluxetable}

\end{document}